\documentclass[12pt]{article}
\usepackage{epsfig}
\usepackage{amssymb}
\usepackage{amsmath,amsfonts,amssymb,graphicx}
\newcommand {\be}{\begin{equation}}
\newcommand {\ee}{\end {equation}}
\newcommand{\beq}{\begin{eqnarray}}
\newcommand{\eeq}{\end{eqnarray}}

\begin{document}
\title{Neutrino Oscillations With Three Active and Three Sterile Neutrinos}
\author{Leonard S. Kisslinger\\
Department of Physics, Carnegie Mellon University, Pittsburgh PA 15213}
\maketitle
\date{}
\noindent
PACS Indices:11.30.Er,14.60.Lm,13.15.+g
\vspace{3mm}


\begin{abstract}
 This is an extension of estimates of the probability  of $\mu$
to $e$ neutrino oscillation with one sterile neutrino to three sterile
neutrinos, using a 6x6 matrix. Since the mixing angle for only one sterile 
neutrino has been experimentally determined, we estimate the $\mu$ to $e$ 
neutrino oscillation probability with different mixing angles for two of the 
sterile neutrinos.
\end{abstract}

\section{Introduction}

Most recent theories of neutrino oscillations have used a 3x3 S-matrix 
approach with three active neutrinos\cite{ahlo01,jo04,hjk11}. Recent 
experiments on neutrino oscillations\cite{mini13}  have suggested the 
existence of at least one sterile neutrino with the mass and mixing angles 
used in the present work. See Ref\cite{mini13} for references to earlier 
experiments, and Refs\cite{kmms13,ggllz15} for reviews of sterile 
neutrino oscillations with references to experimental and theoretical 
publications.

In the present work we use a U-matrix approach, introduced for active
neutrinos with a 3x3 U-matrix\cite{as97}, and extended to a 4x4 U-matrix
with one sterile neutrino in a recent study of $\mathcal{P}(\nu_\mu
\rightarrow \nu_e)$, the transition probability for a muon
neutrino to oscillate to an electron neutrino\cite{lsk14,lsk15}.
We introduce a 6x6 U-matrix for three active and three
sterile neutrinos, an extension of previous work with six neutrinos\cite{tg07}.
An early study of the effect of adding 3 sterile neutrinos may be found in
Ref\cite{gs81}, where it was found that in a broad class of theories consistent
with grand unification, the neutrino mixing angles are likely to be 
comparable to the corresponding quark mixing angles and might be much larger
in a special case. This result holds for a wide range of mass ratios for the 
light-neutrino Majorana masses.
 
\section{ 6x6 U-Matrix}

Active neutrinos with flavors $\nu_e,\nu_\mu,\nu_\tau$ and three sterile 
neutrinos, $\nu_{s_1},\nu_{s_2},\nu_{s_3}$ are related to neutrinos with 
definite mass by
\beq
\label{f-mrelation}
      \nu_f &=& U\nu_m \; ,
\eeq
where $U$ is a 6x6 matrix and $\nu_f,\nu_m$ are 6x1 column vectors.
We use the notation  $s_{ij}, c_{ij}=sin\theta_{ij},cos\theta_{ij}$, with 
$\theta_{12}, \theta_{23}, \theta_{13}$ the mixing angles for active neutrinos; 
and $s_\alpha=sin(\alpha), c_\alpha=cos(\alpha), s_\beta=sin(\beta)$, etc, where 
$\alpha,\beta,\gamma$ are sterile-active neutrino mixing angles. 
\beq
\label{Uform}
     U &=& O^{23}O^{13} O^{12} O^{14} O^{24} O^{34} O^{15} O^{25} O^{35} 
O^{45} O^{16}O^{26} O^{36} O^{46} O^{56} 
\eeq
with ($O^{45}$, $O^{46}$, and $O^{56}$, giving sterile-sterile neutrino
mixing, are not shown)
\vspace{3mm}

$O^{23}$=
 $\left( \begin{array}{ccclcr} 1 & 0 & 0 & 0 & 0 & 0 \\ 0 & c_{23} & s_{23} & 0 
& 0 & 0 \\
0 & -s_{23} & c_{23} & 0 & 0 & 0 \\ 0 & 0 & 0 & 1 & 0 & 0 \\ 0 & 0 & 0 & 0 & 1 &
 0 \\ 0 & 0 & 0 & 0 & 0 & 1 \end{array} \right)$
\hspace{3mm}$O^{13}$=
$\left( \begin{array}{ccclcr} c_{13} & 0 & s_{13} & 0 & 0 & 0 \\ 0 & 1 & 0 & 0 
& 0 & 0 \\-s_{13} & 0  & c_{13} & 0 & 0 & 0 \\ 0 & 0 & 0 & 1 & 0 & 0\\ 0 & 0 & 
0 & 0 & 1 & 0\\  0 & 0 & 0 & 0 & 0 & 1  \end{array} \right)$
\vspace{3mm}

$O^{12}$=
 $\left( \begin{array}{ccclcr} c_{12} & s_{12} & 0 & 0 & 0 & 0\\ -s_{12} & 
c_{12} & 0 & 0 & 0 & 0 \\ 0 & 0  & 1 & 0 & 0 & 0 \\ 0 & 0 & 0 & 1 & 0 & 0\\
0 & 0 & 0 & 0 & 1 & 0\\ 0 & 0 & 0 & 0 & 0 & 1 \end{array} \right)$
\hspace{3mm}$O^{14}$=
 $\left( \begin{array}{ccclcr} c_\alpha & 0 & 0 & s_\alpha & 0 & 0\\ 
 0 & 1  & 0 & 0 & 0 & 0\\  0 & 0 & 1 & 0 & 0 & 0\\
 -s_\alpha & 0 & 0 & c_\alpha & 0 & 0\\
0& 0 & 0 & 0 & 1 & 0 \\ 0 & 0 & 0 & 0 & 0 & 1 \end{array} \right)$
\vspace{3mm}

$O^{24}$=
$ \left( \begin{array}{ccclcr} 1 & 0 & 0 & 0 & 0 & 0 \\ 0 & c_\alpha & 0 & 
s_\alpha & 0 & 0 \\ 0 & 0 & 1 & 0 & 0 & 0 \\ 0 & -s_\alpha & 0 & c_\alpha & 0 & 0 
 \\
 0 & 0  & 0 & 1 & 0  & 0\\
0& 0 & 0 & 0 & 1 & 0 \\ 0 & 0 & 0 & 0 & 0 & 1  \end{array} \right)$
\hspace{3mm}$O^{34}$=
$\left( \begin{array}{ccclcr} 1 & 0 & 0 & 0 & 0 & 0 \\ 0 & 1 & 0 & 0 & 0 & 0 \\
 0 & 0  & c_\alpha & s_\alpha & 0 & 0 \\ 0 & 0 & -s_\alpha & c_\alpha & 0 & 0\\
0& 0 & 0 & 0 & 1 & 0 \\ 0 & 0 & 0 & 0 & 0 & 1   \end{array} \right)$

$O^{15}$=
$\left( \begin{array}{ccclcr} c_\beta & 0 & 0 & 0 & s_\beta & 0\\
 0 & 1 & 0 & 0 & 0 & 0 \\ 0 & 0 & 1 & 0 & 0 & 0 \\ 0 & 0 & 0 & 1 & 0 & 0 \\ 
- s_\beta & 0 & 0 & 0 & c_\beta & 0 \\ 0 & 0 & 0 & 0 & 0 & 1  \end{array} 
\right)$
\hspace{3mm}$O^{25}$=
 $\left( \begin{array}{ccclcr}  1 & 0 & 0 & 0 & 0 & 0 \\ 0 & c_\beta & 0 & 0 &
 s_\beta & 0\\ 0 & 0 & 1 & 0 & 0 & 0 \\ 0 & 0 & 0 & 1 & 0 & 0 \\ 
 0 &-s_\beta  & 0 & 0 & c_\beta & 0 \\ 0 & 0 & 0 & 0 & 0 & 1  \end{array}
\right)$
\vspace{3mm}

$O^{35}$=
$\left( \begin{array}{ccclcr} 1 & 0 & 0 & 0 & 0 & 0 \\ 0 & 1 & 0 & 0 & 0 & 0 \\
 0 & 0 & c_\beta & 0 & s_\beta & 0 \\ 0 & 0 & 0 & 1 & 0 & 0 \\ 0& 0 
 & -s_\beta  & 0 & c_\beta & 0 \\ 0 & 0 & 0 & 0 & 0 & 1 \end{array} \right)$
\hspace{3mm}$O^{16}$=
 $\left( \begin{array}{ccclcr} c_\gamma & 0 & 0 & 0  & 0 & s_\gamma\\
 0 & 1 & 0 & 0 & 0 & 0 \\ 0 & 0 & 1 & 0 & 0 & 0 \\ 0 & 0 & 0 & 1 & 0 & 0 \\
 0 & 0 & 0 & 0 & 1 & 0 \\ - s_\gamma & 0 & 0 & 0 & 0 & c_\gamma   \end{array} 
\right)$
\vspace{3mm}

$O^{26}$=
 $\left( \begin{array}{ccclcr}  1 & 0 & 0 & 0 & 0 & 0 \\ 0 & c_\gamma & 0 & 0 
 & 0 & s_\gamma\\ 0 & 0 & 1 & 0 & 0 & 0 \\ 0 & 0 & 0 & 1 & 0 & 0 \\ 
 0 & 0 & 0 & 0 & 1 & 0 \\ 
 0 &-s_\gamma  & 0 & 0 & 0 & c_\gamma  \end{array}
\right)$
\hspace{3mm}$O^{36}$=
 $\left( \begin{array}{ccclcr} 1 & 0 & 0 & 0 & 0 & 0 \\ 0 & 1 & 0 & 0 & 0 & 0 \\
 0 & 0 & c_\gamma & 0  & 0 & s_\gamma\\
 0  & 0 & 0 & 1 & 0 & 0 \\ 0 & 0 & 0 & 0 & 1 & 0 \\  0 & 0 &- s_\gamma & 0 & 0
 & c_\gamma   \end{array} 
\right)$
\vspace{5mm}

  $ \mathcal{P}(\nu_\mu \rightarrow\nu_e)$ 
is obtained from the 6x6 U matrix and the neutrino mass differences
$\delta m_{ij}^2=m_i^2-m_j^2$ for a neutrino beam with energy $E$ and baseline
$L$ by
\beq
\label{Pue-1}
 \mathcal{P}(\nu_\mu \rightarrow\nu_e) &=& Re[\sum_{i=1}^{6}\sum_{j=1}^{6}
U_{1i}U^*_{1j}U^*_{2i}U_{2j} e^{-i(\delta m_{ij}^2/E)L}] \; ,
\eeq
an extension the 4x4\cite{lsk14,lsk15} theory with one serile neutrino, 
which used the  3x3 formalism of Ref\cite{as97}, to a 6x6 matrix 
formalism\cite{tg07}.
From Eq(\ref{Uform}), multiplying the 12 6x6 $O$ matrices, we obtain the 
matrix U. With $\delta_{CP}$=0, $U^*_{ij}=U_{ij}$, so we only need $U_{1j},U_{2j}$.

\beq
\label{U1j}
  U_{11}&=&.821 ca{\rm \;}cb{\rm \;}cg  \nonumber \\
  U_{12} &=& cg ((.554 ca - .821 sa^2) cb - .821 ca{\rm \;}sb^2) 
- .821 ca{\rm \;} cb{\rm \;}sg^2 \nonumber \\
 U_{13}&=&cg ((.15 ca-.554 sa^2-.821ca{\rm \;}sa^2)cb-(.554 ca - .821 sa^2)sb^2
\nonumber \\
      && -.821ca{\rm \;}cb{\rm \;}sb^2) - .821 ca{\rm \;}cb{\rm \;}cg{\rm \;}
sg^2 - ((.554 ca - .821 sa^2) cb -.821 ca{\rm \;}sb^2)sg^2 \nonumber \\
 U_{14} &=&cg(cb(.15sa +.554 ca{\rm \;}sa + .821 ca^2{\rm \;}sa)-.821ca{\rm \;}
cb^2{\rm \;}sb^2
\nonumber \\
   && -(.554 ca-.821 sa^2)cb{\rm \;}sb^2-(.15 ca-.554 sa^2-.821 ca sa^2)sb^2)
 - .821ca{\rm \;}cb{\rm \;}sg^2 cg^2\nonumber \\
  &&-cg((.554ca-.821sa^2) cb -.821ca {\rm \;}sb^2)sg^2
-(cb (.15 ca - .554 sa^2 - .821 ca sa^2)\nonumber \\
  && - .821ca{\rm \;}cb{\rm \;}sb^2 -(.554 ca-.821 sa^2) sb^2)sg^2 
\nonumber 
\eeq
\newpage
\beq
\label{U1j2}
U_{15} &=&cg(.821ca{\rm \;}sb{\rm \;}cb^3+(.15sa+.554ca{\rm \;}sa+
.821 ca^2{\rm \;}sa)sb \nonumber \\
   && +(.554 ca-.821 sa^2)cb^2{\rm \;}sb+(.15ca-.554sa^2-.821ca{\rm \;}sa^2)
cb{\rm \;}sb)
\nonumber \\
 && -.821ca{\rm \;}cb{\rm \;}cg^3 sg^2 -cg^2(cb(.554ca-.821 sa^2)-.821 sb^2)sg^2
\nonumber \\
  &&-cg(cb (.15 ca-.554 sa^2-.821ca{\rm \;}sa^2)-.821 ca{\rm \;}cb{\rm \;}sb^2
 \nonumber \\ 
   && -(.554 ca-.821 sa^2)sb^2 sg^2 -(cb(.15sa+.554ca{\rm \;} sa +
.821ca^2 sa) - .821 ca{\rm \;}cb^2{\rm \;}sb^2  \nonumber \\
   && -cb(.554ca -.821sa^2)sb^2+(.15ca-.554sa^2-.821ca{\rm \;}sa^2)sb^2) sg^2
 \nonumber \\ 
 U_{16}&=& .821ca{\rm \;}cb{\rm \;}sg{\rm \;}cg^4+(.821ca{\rm \;}cb^3{\rm \;}sb
 + (.15sa+.554ca{\rm \;}sa+.821ca^2{\rm \;}sa)sb\nonumber \\
   && +cb^2 (.554 ca-.821 sa^2)sb + 
    cb(.15 ca -.554 sa^2-.821 ca{\rm \;} sa^2) sb) sg \nonumber \\
   &&+cg^3 ((.554 ca - .821 sa^2) cb - .821 ca{\rm \;}sb^2)sg +\nonumber \\
   && cg^2 (cb (.15 ca -.554 sa^2 - .821ca{\rm \;}sa^2)-.821ca{\rm \;}cb
{\rm \;}sb^2\nonumber \\
   && - (.554 ca - .821 sa^2) sb^2) sg\nonumber \\
   &&+cg(cb(.15 sa+.554ca{\rm \;}sa+.821 ca^2{\rm \;}sa)-.821ca{\rm \;}cb^2 sb^2
\nonumber \\
  && -cb(.554ca-.821sa^2)sb^2-(.15 ca-.554 sa^2-.821ca{\rm \;}sa^2)sb^2) sg
\; ,
\eeq

\beq
\label{U2j}
  U_{21}&=& -.484ca{\rm \;}cb{\rm \;}cg \nonumber \\
  U_{22}&=&cg(.527ca+.484 sa^2)cb-.821ca{\rm \;}sb^2)+
.484ca{\rm \;}cb{\rm \;}sg^2
\nonumber \\
  U_{23}&=& cg((.699ca-.527sa^2+.484ca{\rm \;}sa^2)cb-(.527ca+.484sa^2)sb^2 
+.484ca{\rm \;}cb{\rm \;}sb^2)\nonumber \\
  && +.484ca{\rm \;}cb{\rm \;}cg{\rm \;}sg^2-((.527ca + .484sa^2)cb
+.484 ca{\rm \;}sb^2)*sg^2 
\nonumber \\
  U_{24}&=& cg(cb(.699 sa+.527ca{\rm \;}sa-.484ca^2{\rm \;}sa)+
.484ca{\rm \;}cb^2{\rm \;}sb^2 \nonumber \\
   && -(.527ca +.484sa^2)cb{\rm \;}sb^2 -(.699ca-.527sa^2+.484ca{\rm \;}sa^2) 
sb^2) +.484ca{\rm \;}cb{\rm \;}sg^2{\rm \;}cg^2 \nonumber \\
   &&-cg((.527 ca +.484sa^2)cb+.484ca{\rm \;}sb^2)sg^2-(cb(.69 ca-.527sa^2+
 .484ca{\rm \;}sa^2) +\nonumber \\ 
   && .484ca{\rm \;}cb{\rm \;}sb^2-(.527ca + .484 sa^2)sb^2)sg^2 
 \nonumber \\
  U_{25}&=& cg(-.484 ca{\rm \;}sb{\rm \;}cb^3 +(.699sa+.527ca{\rm \;}sa
-.484ca^2{\rm \;}sa)sb\nonumber \\
   && +(.527ca+.484sa^2)cb^2{\rm \;}sb +(.699ca -.527sa^2+.484 ca{\rm \;}sa^2) 
cb{\rm \;}sb)+.484ca{\rm \;}cb{\rm \;}cg^3{\rm \;}sg^2 \nonumber \\
   &&-cg^2(cb(.527ca+.484sa^2)+.484ca{\rm \;}sb^2)sg^2
-cg(cb(.699ca-.527sa^2+.484ca{\rm \;}sa^2)+
\nonumber \\
    && .484ca{\rm \;}cb{\rm \;}sb^2-(.527ca+.484sa^2)sb^2)sg^2
-(cb(.699sa+.527ca{\rm \;}sa-.484ca^2{\rm \;}sa)+ \nonumber \\
   &&.484ca{\rm \;}cb^2{\rm \;}sb^2 -cb(.527ca +.484sa^2)sb^2
+(.699ca-.527sa^2+.484 ca{\rm \;}sa^2)sb^2)sg^2 \nonumber \\
 U_{26}&=& -.484ca{\rm \;}cb{\rm \;}sg{\rm \;}cg^4+
(-.484ca{\rm \;}cb^3{\rm \;}sb + (.699 sa + .527 ca sa-.484ca^2{\rm \;}sa)sb
\nonumber \\
   && +cb^2 (.527ca + .484sa^2)sb+ cb(.699ca-.527sa^2+.484ca{\rm \;}sa^2)sb)sg
\nonumber\\
   && +cg^3((.527ca+.484 sa^2)cb + .484 ca{\rm \;}sb^2)sg +\nonumber\\
   && cg^2(cb(.699ca-.527sa^2+.484ca{\rm \;}sa^2)+.484 ca{\rm \;}cb{\rm \;}sb^2
    - (.527 ca + .484 sa^2) sb^2) sg \nonumber\\
   &&+cg(cb(.699sa+.527ca{\rm \;}sa-.484ca^2 sa)+.484ca{\rm \;}cb^2{\rm \;}sb^2
   - cb(.527ca +.484sa^2)sb^2 \nonumber\\
   &&-(.699 ca-.527sa^2 +.484 ca{\rm \;}sa^2)sb^2)sg \; .
\eeq
\newpage

\section{$\mathcal{P}(\nu_\mu \rightarrow \nu_e)$ For equal sterile
 neutrino masses}

Assuming that all three sterile neutrinos have the same mass, sterile-active
neutrino mass differences are $\delta m_{4j}^2=m_4^2-m_j^2 \simeq .9 (eV)^2$,
with $\delta m_{4j}^2$ taken from the best fit to neutrino oscillation 
data\cite{mini13} (see Ref\cite{mini13} for references to earlier experiments),
from Eq(\ref{Pue-1}) $\mathcal{P}(\nu_\mu \rightarrow \nu_e)$ is 
\beq
\label{Pue-2}
\mathcal{P}(\nu_\mu \rightarrow \nu_{e}) &=&Re[U_{11}U_{21}[ U_{11}U_{21}+
 U_{12}U_{22} e^{-i\delta L}+ U_{13}U_{23} e^{-i\Delta L}+ \nonumber \\
  && (U_{14}U_{24}+U_{15}U_{25} +U_{16}U_{26}) e^{-i\gamma L}]+ \nonumber \\
  &&  U_{12}U_{22}[ U_{11}U_{21}e^{-i\delta L}+ U_{12}U_{22} + U_{13}U_{23} 
e^{-i\Delta L}+ \nonumber\\
  && (U_{14}U_{24}+U_{15}U_{25} +U_{16}U_{26}) e^{-i\gamma L}]+ \nonumber \\
  &&  U_{13}U_{23}[ U_{11}U_{21}e^{-i\Delta L}+ U_{12}U_{22}e^{-i\Delta L}
 \nonumber \\
  && + U_{13}U_{23} 
    +(U_{14}U_{24}+U_{15}U_{25} +U_{16}U_{26}) e^{-i\gamma L}]+ \nonumber \\
   &&  (U_{14}U_{24}+U_{15}U_{25}+U_{16}U_{26})[(U_{11}U_{21}+ U_{12}U_{22}
  \nonumber \\
   &&+ U_{13}U_{23})e^{-i\gamma L}+U_{14}U_{24}+U_{15}U_{25}+U_{16}U_{26}]] \; ,
\eeq 
with $\delta=\delta m_{12}^2/2E,\; \Delta=\delta m_{13}^2/2E,\; \gamma=
\delta m_{jk}^2/2E$ (j=1,2,3;k=4,5,6). The neutrino mass differences are 
$\delta m_{12}^2=7.6 \times 10^{-5}(eV)^2$, $\delta m_{13}^2 = 2.4\times 10^{-3} 
(eV)^2$; and  $\delta m_{jk}^2 (j=1,2,3;k=4,5,6) =0.9 (eV)^2$\cite{mini13}.
\vspace{3mm}

 From Eq(\ref{Pue-2})
\beq
\label{Pue-3}
 \mathcal{P}(\nu_\mu \rightarrow\nu_e) &=& U_{11}^2 U_{21}^2+
 U_{12}^2 U_{22}^2+ U_{13}^2 U_{23}^2+  \nonumber \\
  && (U_{14}U_{24}+U_{15}U_{25}+U_{16}U_{26})^2 + \nonumber \\
  &&  2U_{11} U_{21} U_{12} U_{22} cos\delta L + \\
  && 2(U_{11} U_{21} U_{13} U_{23}+ U_{12} U_{22} U_{13} U_{23})cos\Delta L+
\nonumber \\
  &&2(U_{14}U_{24}+U_{15}U_{25}+U_{16}U_{26}) \nonumber \\
  &&(U_{11} U_{21}+U_{12} U_{22}+U_{13} U_{23})cos\gamma L \nonumber \; .
\eeq 
\vspace{3mm}

  Note that $\alpha\simeq 9.2^o$ from a recent analysis of MiniBooNE data, 
which was used in a recent study of $\mathcal{P}(\nu_\mu \rightarrow\nu_e)$ 
with one sterile neutrino\cite{lsk14,lsk15}. The figure below shows 
$\mathcal{P}(\nu_\mu \rightarrow \nu_e)$ with $\alpha=\beta=\gamma= 0^o$,
giving the results of a recent 3x3 S-mtrix calculation\cite{lsk14-2}.
In the figure, for the other curves, the sterile-active mixing angle 
$\alpha=9.2^o$, while $\beta$ and  $\gamma$ are chosen to be $9.2^o$ and 
$20^o$ to compare the 6x6 to the previous 3x3 results. 

\newpage
  Using Eq(\ref{Pue-3}), one finds $\mathcal{P}(\nu_\mu \rightarrow \nu_e)$ for 
the 6x6 vs 3x3 theories:

\vspace{5cm}

\begin{figure}[ht]
\begin{center}
\epsfig{file=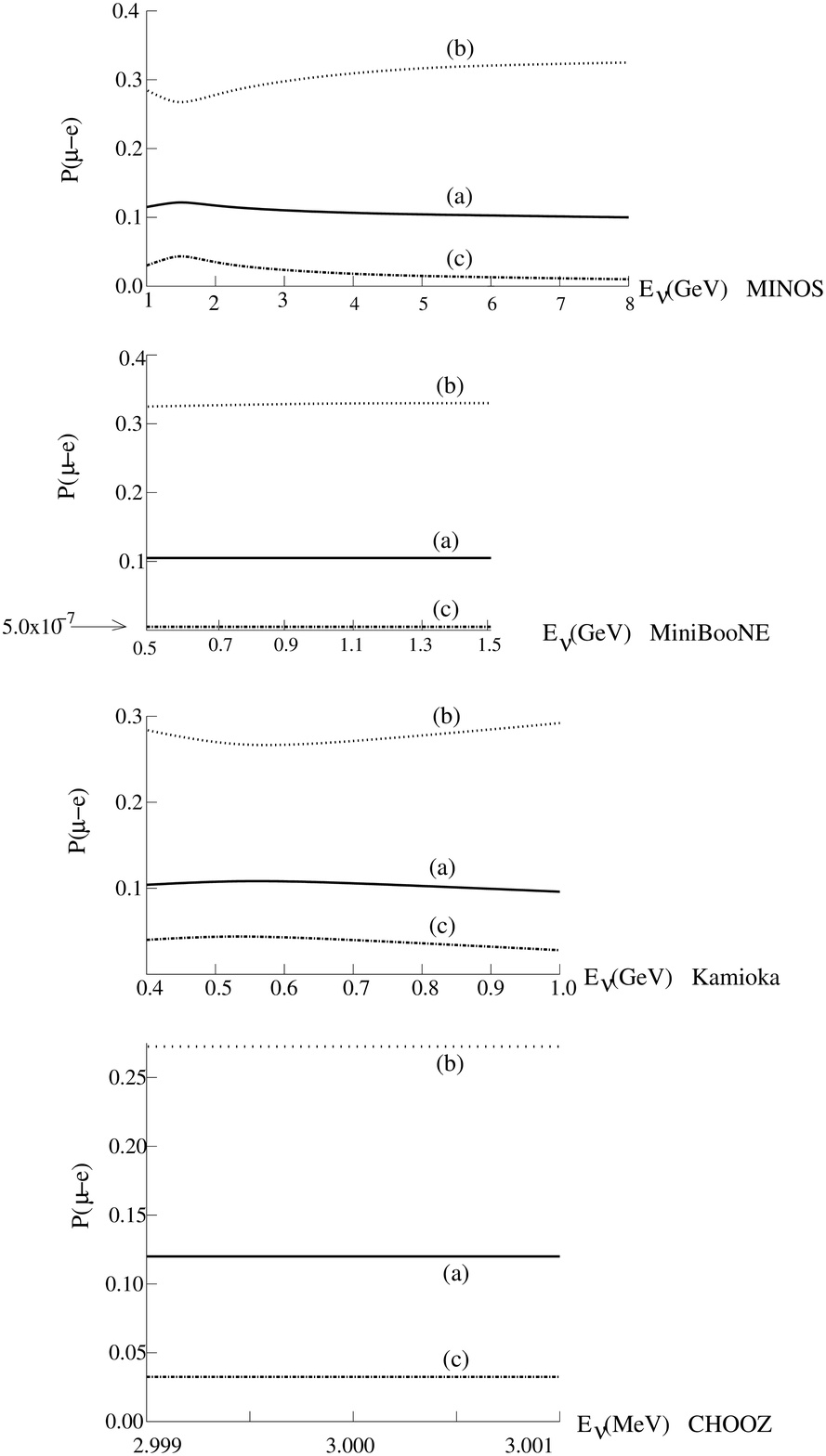,height=12cm,width=10cm}
\end{center}
\caption{$\mathcal{P}(\nu_\mu \rightarrow\nu_e)$ 
for MINOS(L=735 km), MiniBooNE(L=500m), JHF-Kamioka(L=295 km), and 
CHOOZ(L=1.03 km). (a) solid, for $\alpha=\beta=\gamma$=$9.2^o$;  
(b) dashed, for $\alpha, \beta, \gamma$ =$9.2^o$, $20^o$, $20^o$; 
(c) dash-dotted curve for $\alpha=\beta=\gamma$=$0^o$ giving the 3x3 result .}
\end{figure}
\vspace{3mm}
\newpage
\section{Conclusions}

  From the figure we note that with the small mixing angle, $\alpha=\beta=
\gamma$=$9.2^o$, taken from the MiniBooNE analysis for $s_\alpha$, for MINOS,
MiniBooNE, and JHF-Kamioka there is significant difference between our 6x6 and 
the earlier 3x3 prediction for $\mathcal{P}(\nu_\mu \rightarrow \nu_e)$, 
given by  $\alpha=\beta=\gamma=0^o$. For the larger $20^o$ mixing angles for 
$\beta$ and $\gamma$, which are not known, there is a much larger difference 
between th 6x6 and 3x3 theories for these three experimental set-ups.
For CHOOZE, however, $\mathcal{P}(\nu_\mu \rightarrow \nu_e)$ is not 
significantly dependent on the mixing angles  $\alpha, \beta, \gamma$ for
the values used, and is similar to the 3x3 prediction.

 There are many different choices for the parameters needed for this study, 
which we shall investgate in future work.

\Large
{\bf Acknowledgements}
\vspace{3mm}

\normalsize
This work was carried out while LSK was a visitor at Los Alamos
National Laboratory, Group P25. The author thanks Dr. William Louis for 
information about recent and future neutrino oscillation experiments,
and Dr. Terrance Goldman for advice on the mixing angles.

\end{document}